\documentclass{llncs}
\usepackage{amsmath}
\usepackage{svg}
\usepackage{amssymb}
\usepackage{lipsum}
\usepackage{listings}
\usepackage{gensymb}
\usepackage{fancyhdr}
\usepackage{graphicx}
\usepackage{url}
\usepackage{lineno}
\usepackage{xcolor}
\usepackage{color}
\usepackage{graphicx}
\usepackage{geometry}
\newcommand{\fracnot}[3]{\mathcal{F}_{#1}^{#2,#3}}

\begin{document}
\title{Accelerating Compact Fractals with Tensor Core GPUs}
\subtitle{**Technical Report, October 22th 2021**} 

\author{Felipe A. Quezada\inst{1} \and Crist\'obal A. Navarro\inst{1}}

\institute{Temporal Research Group, Informatics Institute, Universidad Austral de Chile \\
\email{cnavarro@inf.uach.cl}}

\maketitle              
\begin{abstract}

This work presents a GPU thread mapping approach that allows doing fast parallel stencil-like computations on discrete fractals using their compact representation. The intuition behind is to employ two GPU tensor-core accelerated thread maps, $\lambda(\omega)$ and $\nu(\omega)$, which act as threadspace-to-dataspace and dataspace-to-threadspace functions, respectively. By combining these maps, threads can access compact space and interact with their neighbors. The cost of the maps is $\mathcal{O}(\log \log(n))$ time, with $n$ being the side of a $n \times n$ embedding for the fractal in its expanded form. The technique works on any fractal that belongs to the Non-overlapping-Bounding-Boxes (NBB) class of discrete fractals, and can be extended to three dimensions as well. Results using an A100 GPU on the Sierpinski Triangle as a case study show up to $\sim11\times$ of speedup and a memory usage reduction of  $234\times$ with  respect  to  a Bounding Box approach.  These results show that the proposed compact approach can allow the scientific community to tackle larger problems that did not fit in GPU memory before, and run even faster than a bounding box approach. 

\end{abstract}
\section{Introduction}
\label{sec:intro}
\begin{figure}[ht!]
\centering
\includegraphics[scale=0.25]{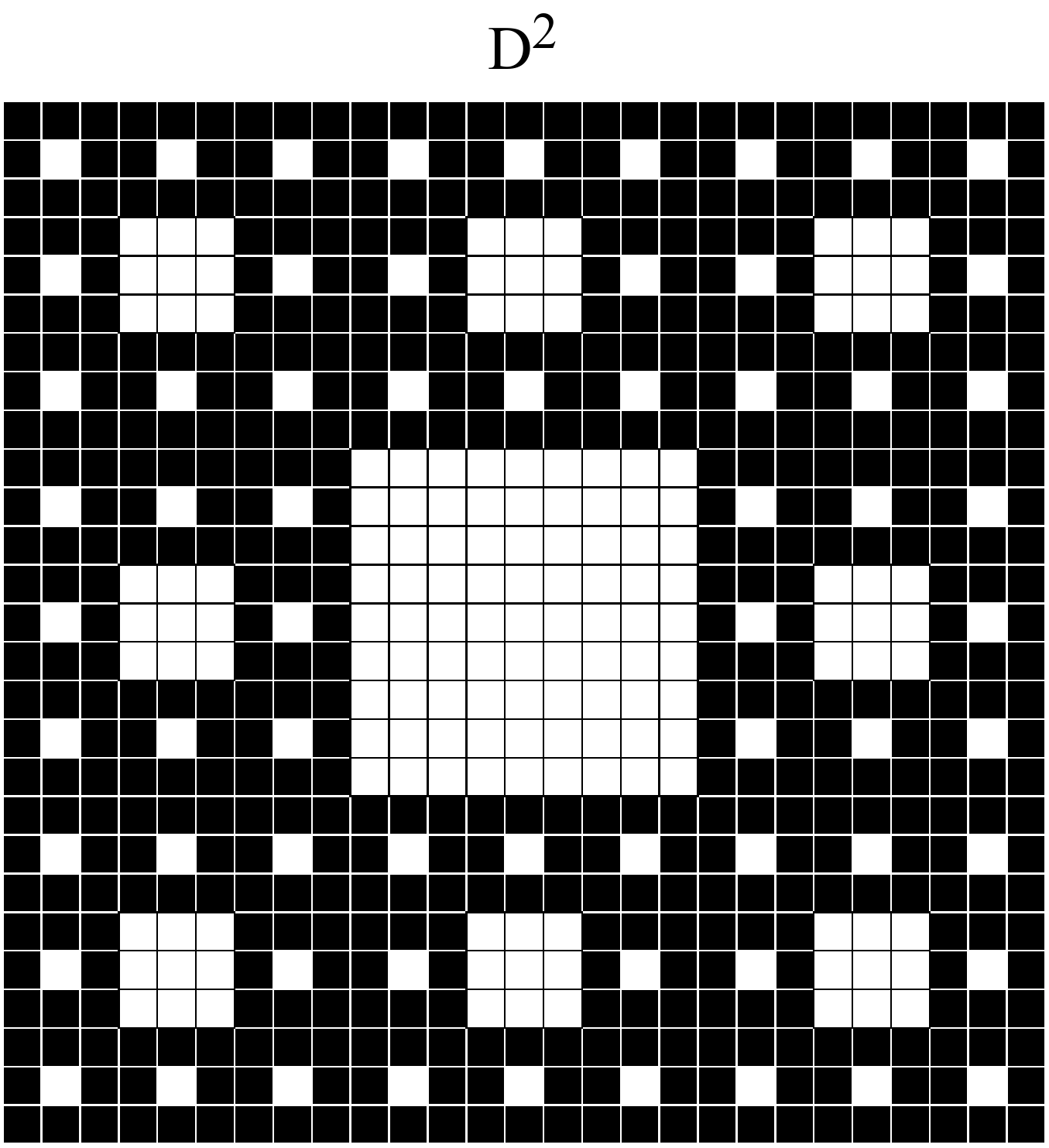}
\caption{The Sierpiński Carpet embedded in a discrete euclidean domain of $27 \times 27$ elements.}
\label{fig:carpet-fractal-embedded}
\end{figure}

Many natural phenomena exhibit fractal properties in their structure, such as vegetation growth \cite{Oppenheimer:1986:RTD:15886.15892,Palmer1988}, terrain formation \cite{MILNE198867}\cite{4767591}, molecular dynamic patterns \cite{rothemund2004}, blood vessels generation \cite{PhysRevLett.90.118101}, among many other examples.
Fractals structures exhibit self-similarity, that is, they manifest the information of the whole structure at different scale levels. Mathematical definitions of fractal geometry have been formulated in order to describe natural phenomena that cannot be easily explained in terms of traditional Euclidean geometry. Parallel computer simulations that act on discrete fractal domains can often use an \textit{embedded} representation, in which the fractal is contained inside a \textit{bounding box} embedding. At first glance, using the embedded representation of the fractal seems attractive as it simplifies the mapping of threads onto data-elements and the exploration of neighbors, at the cost of discarding threads that fall outside the region of interest. Figure \ref{fig:carpet-fractal-embedded} depicts a discrete fractal embedded in a discrete 2D euclidean space.

\begin{figure}[ht!]
\centering
\includegraphics[scale=0.5]{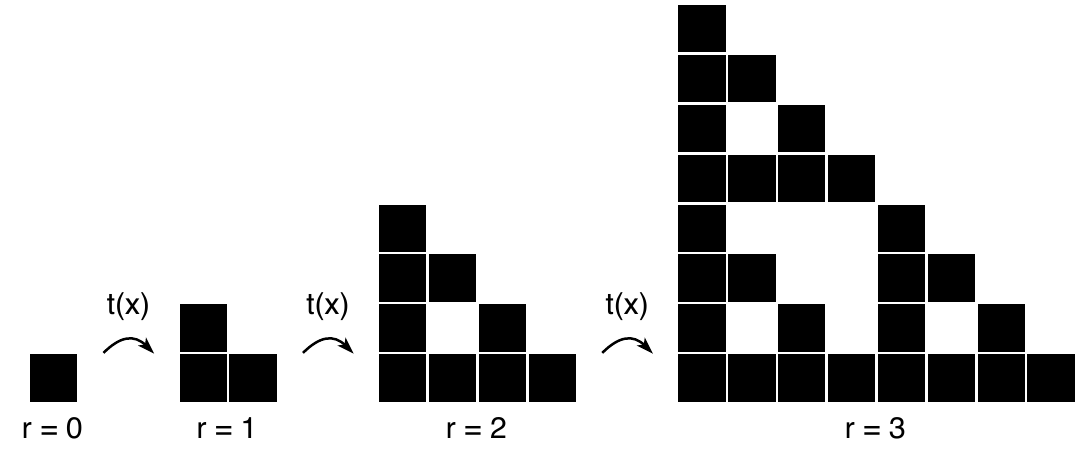}
\caption{The construction of an NBB fractal depicted with the Sierpinski Triangle.}
\label{fig:nbb-construction}
\end{figure}

Navarro \textit{et al.} (2020) studied the Non-overlapping-Bounding-Boxes (NBB) class of discrete fractals, which satisfy two properties: i) the smallest level of the fractal occupies one unit of space and from that point on, it can only scale up, and ii) each fractal has a unique transition function $t(x)$ that takes the current fractal at scale level $r$, and replicates it in the space to generate the fractal at scale level $r+1$. The replicas can be translated, but cannot overlap each other neither rotate. Figure \ref{fig:nbb-construction} shows an example construction of the Sierpinski Triangle.

The notation $\fracnot{n}{k}{s}$ will be used to denote a fractal of the NBB class. Here, $n \in \mathbb{N}$ is the linear size of the fractal along one axis, $k \in \mathbb{N}$ the number of self-similar replicas to generate for the next scale level and $s \in \mathbb{N}$ the growth ratio of $n$ in the next scale level, along an axis. For example, the Sierpiński Carpet (Figure \ref{fig:carpet-fractal-embedded}) is $\fracnot{n}{8}{3}$ and the Sierpiński Triangle (Figure \ref{fig:nbb-construction}) is $\fracnot{n}{3}{2}$. Many different NBB fractals can be described using the same parameters. Table \ref{tbl:discrete-fractals} shows some NBB examples.
\begin{table}[ht!]
    \begin{center}
    \begin{tabular}{ p{2.0cm} | c | c | p{5cm} }
    Fractal Name & Illustration & NBB Step & Hausdorff Dimension ($\mathcal{H} = \frac{\log(k)}{\log(s)}$)\\
    \hline
    Sierpinski Gasket & \raisebox{-\totalheight}{\includegraphics[scale=0.03] {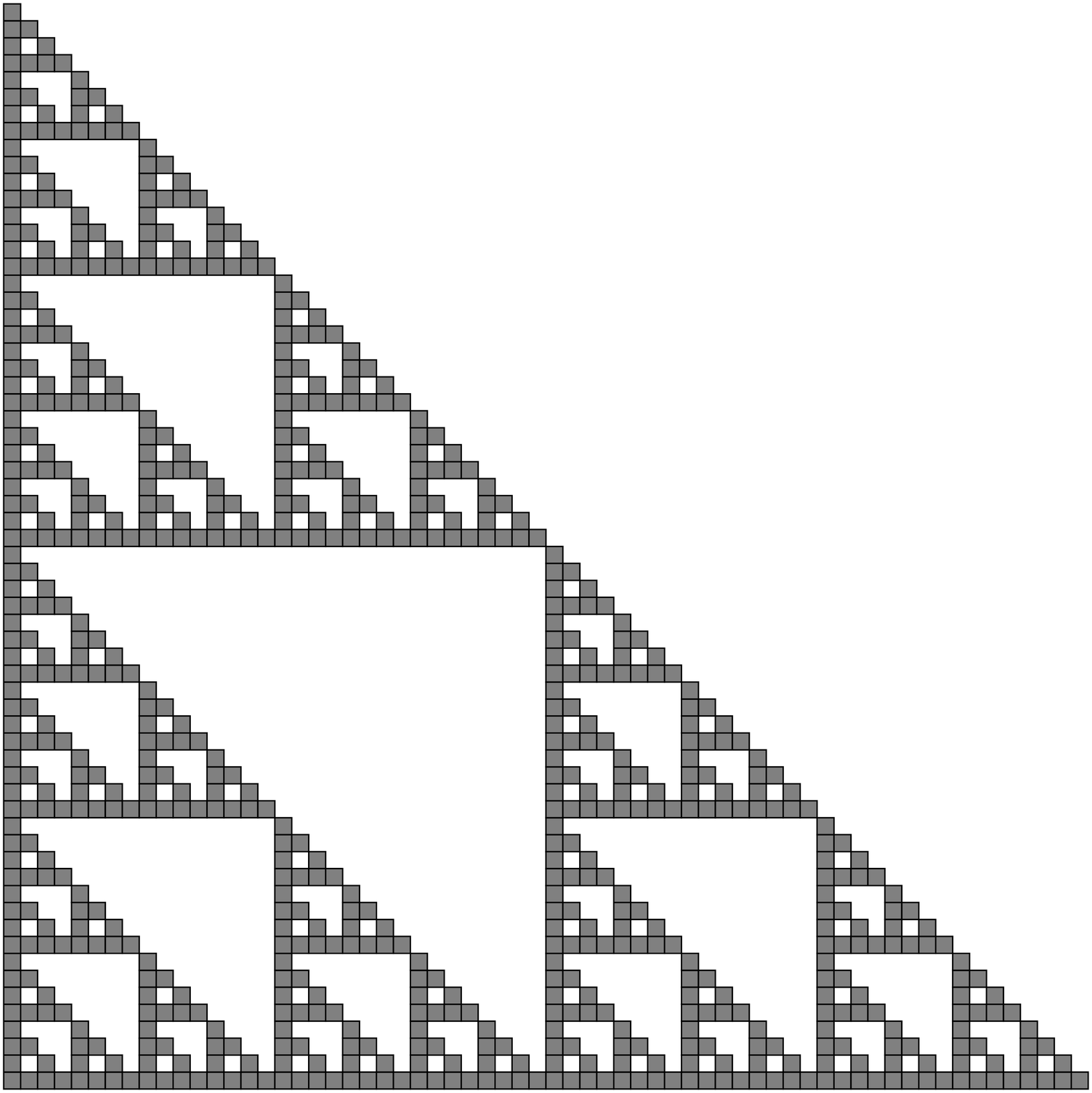}}  & \raisebox{-\totalheight}{\includegraphics[scale=0.5] {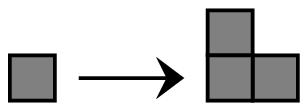}}  & \ \newline $\frac{\log(3)}{\log(2)} \approx 1.58$\\
    \hline
    Chandelier (Custom) & \raisebox{-\totalheight}{\includegraphics[width=2.0cm, height=0.7cm] {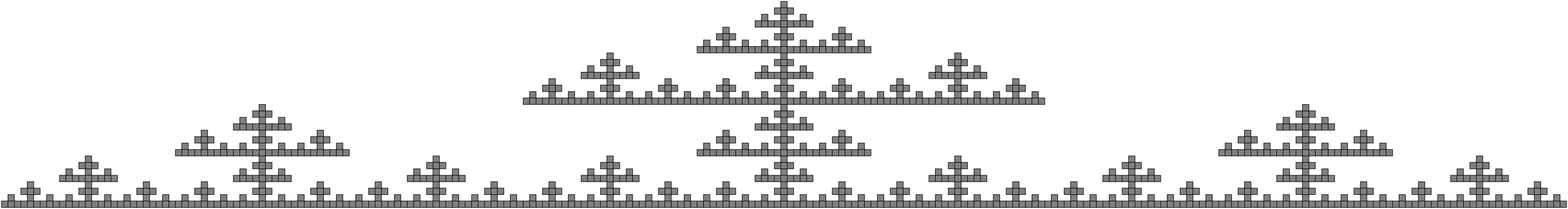}} & \raisebox{-\totalheight}{\includegraphics[scale=0.5] {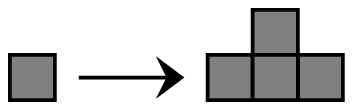}}  & \ \newline $\frac{\log(4)}{\log(3)} \approx 1.26$\\
    \hline
    H-Fractal & \raisebox{-\totalheight}{\includegraphics[scale=0.03] {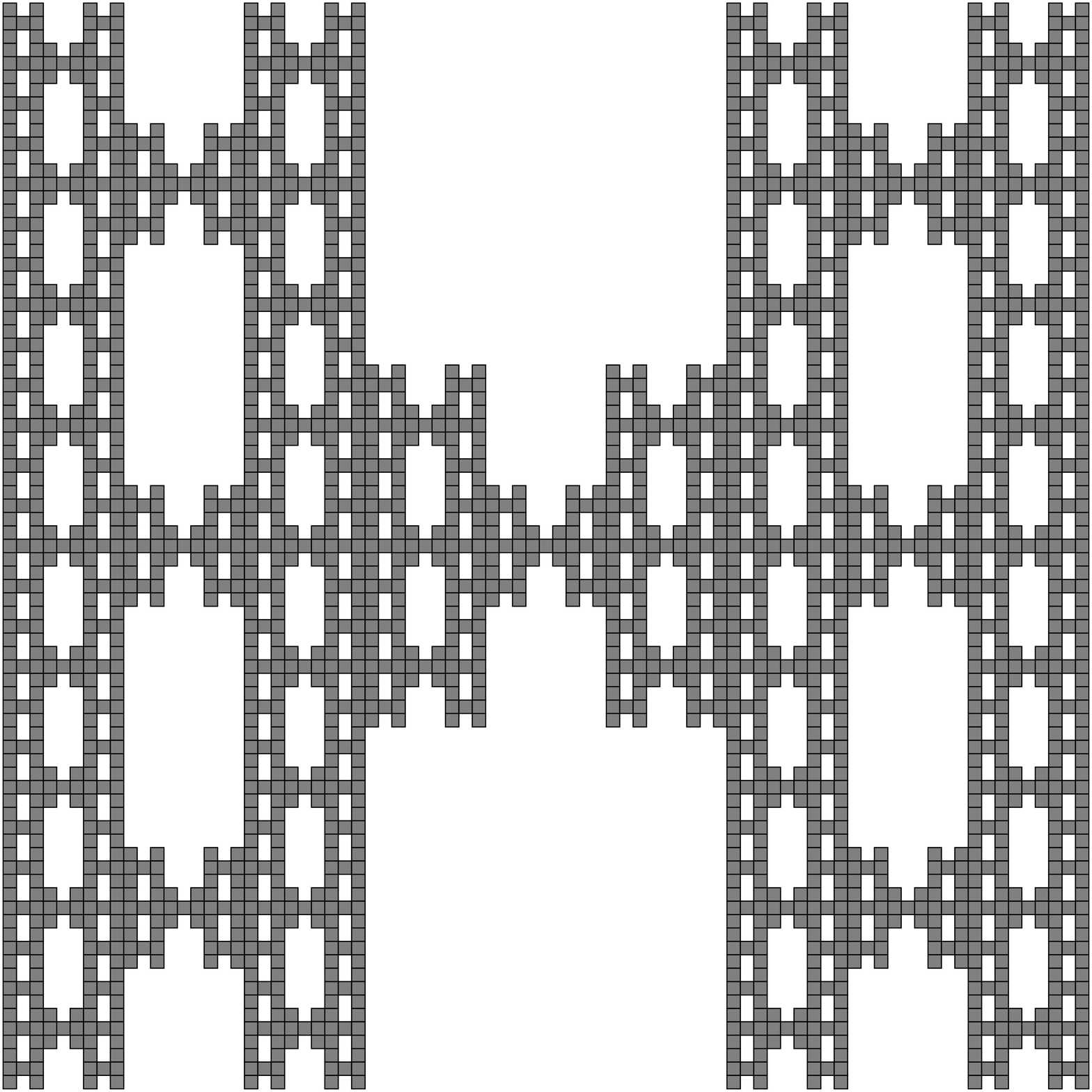}} & \raisebox{-\totalheight}{\includegraphics[scale=0.5] {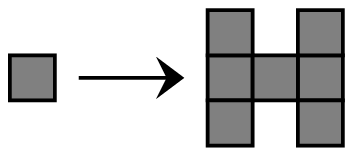}}  & \ \newline $\frac{\log(7)}{\log(3)} \approx 1.77$\\
    \hline
    Candy (Custom) & \raisebox{-\totalheight}{\includegraphics[scale=0.012] {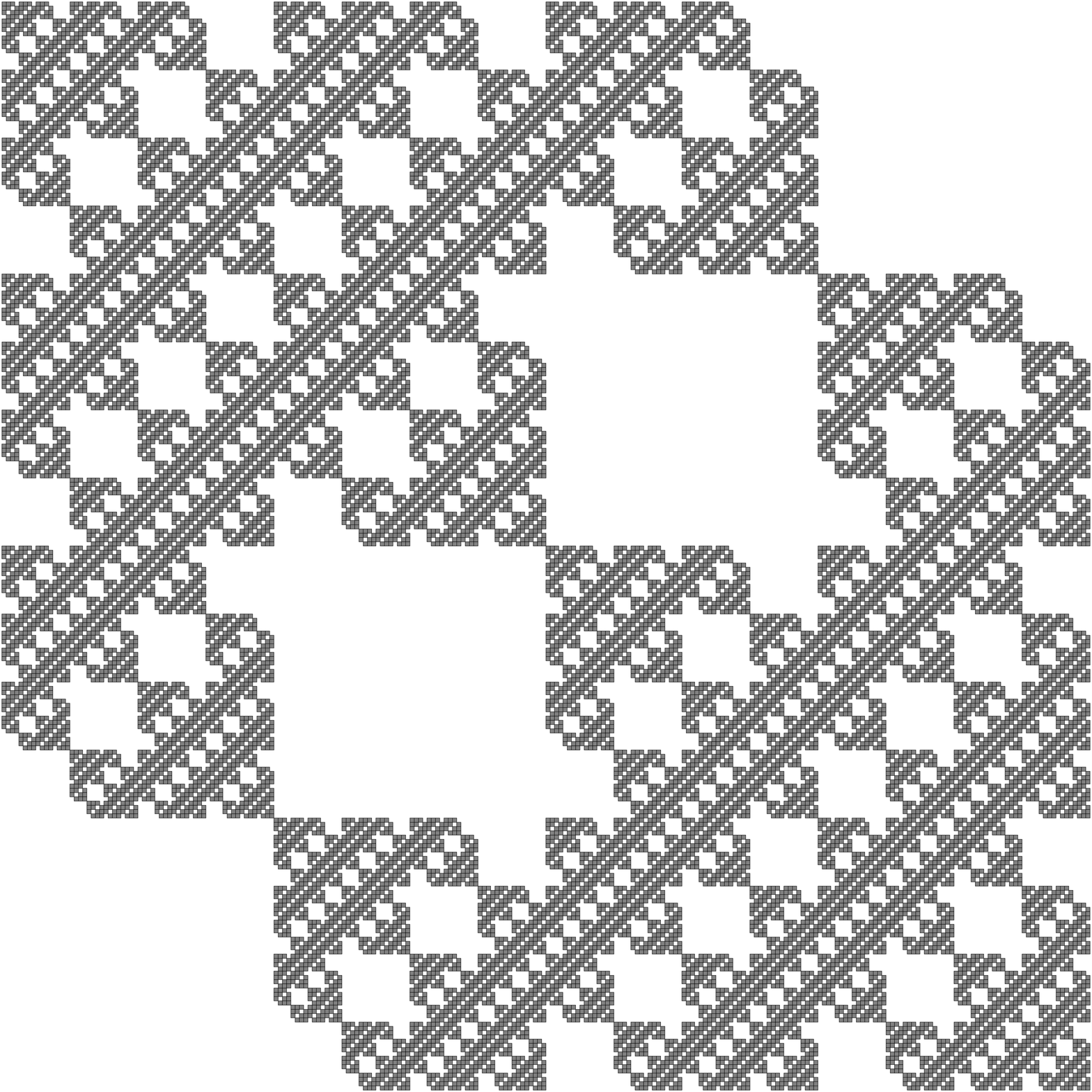}} & \raisebox{-\totalheight}{\includegraphics[scale=0.5] {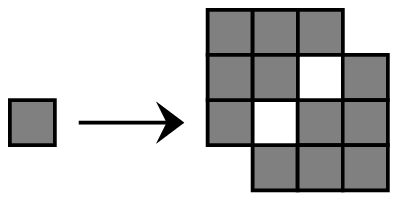}}  & \ \newline $\frac{\log(12)}{\log(4)} \approx 1.79$\\
    \hline
    Sierpinski Carpet & \raisebox{-\totalheight}{\includegraphics[scale=0.035] {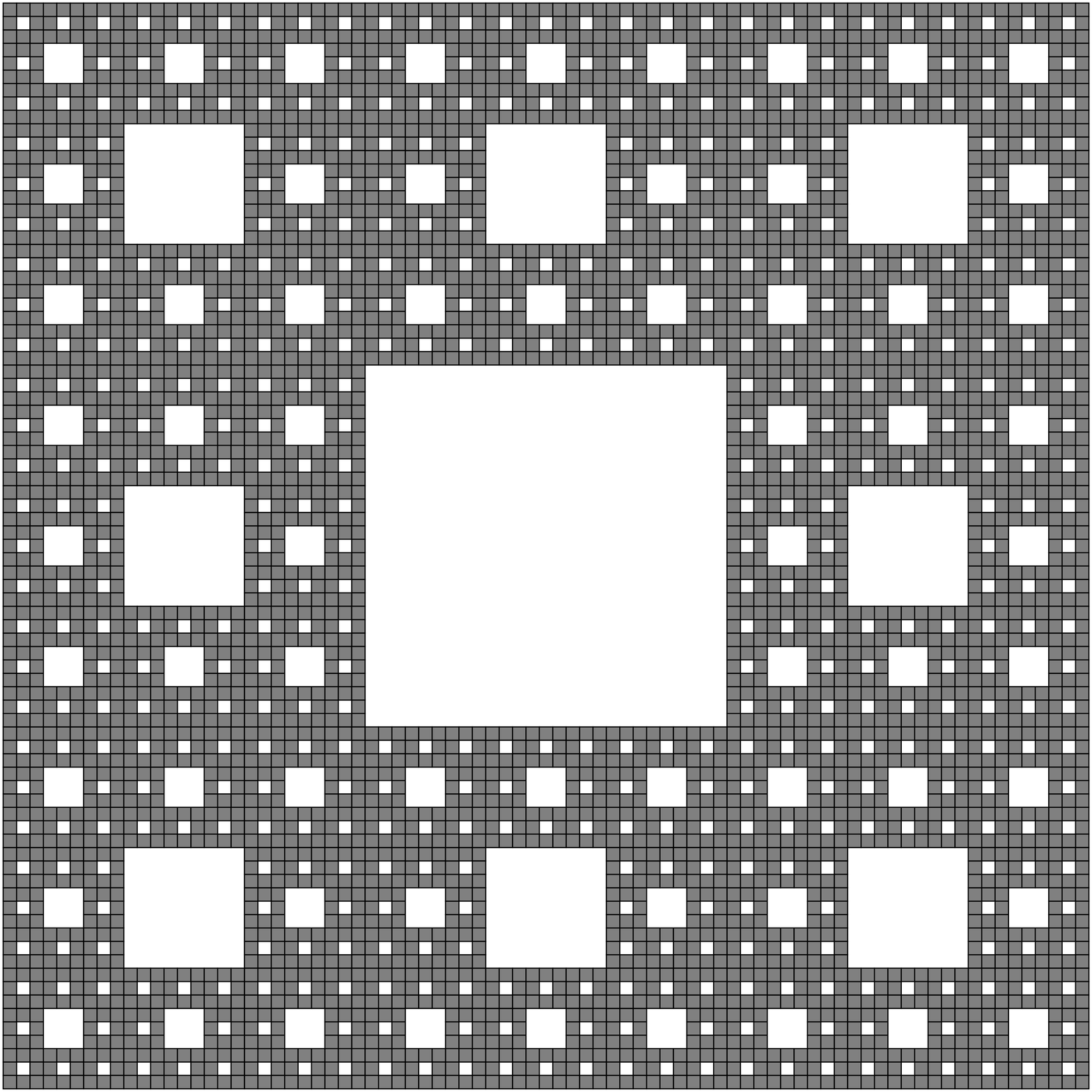}} & \raisebox{-\totalheight}{\includegraphics[scale=0.5] {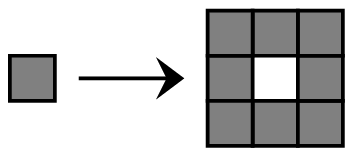}}  & \ \newline $\frac{\log(8)}{\log(3)} \approx 1.89$\\
    \hline
    X-Fractal (Custom) & \raisebox{-\totalheight}{\includegraphics[scale=0.010] {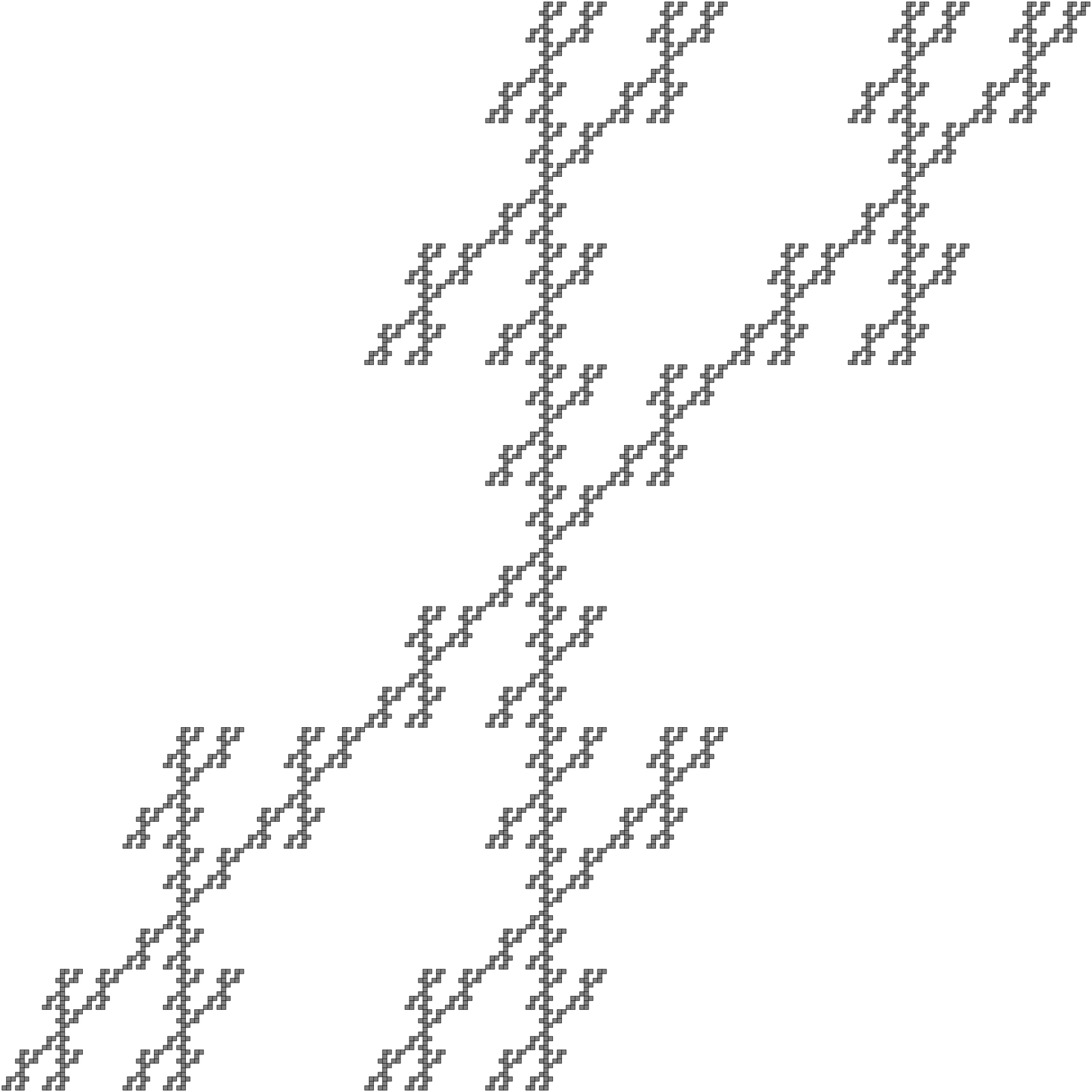}} & \raisebox{-\totalheight}{\includegraphics[scale=0.5] {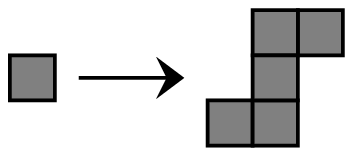}}  & \ \newline $\frac{\log(5)}{\log(3)} \approx 1.46$\\
    \hline
    Vicsek Fractal & \raisebox{-\totalheight}{\includegraphics[scale=0.010] {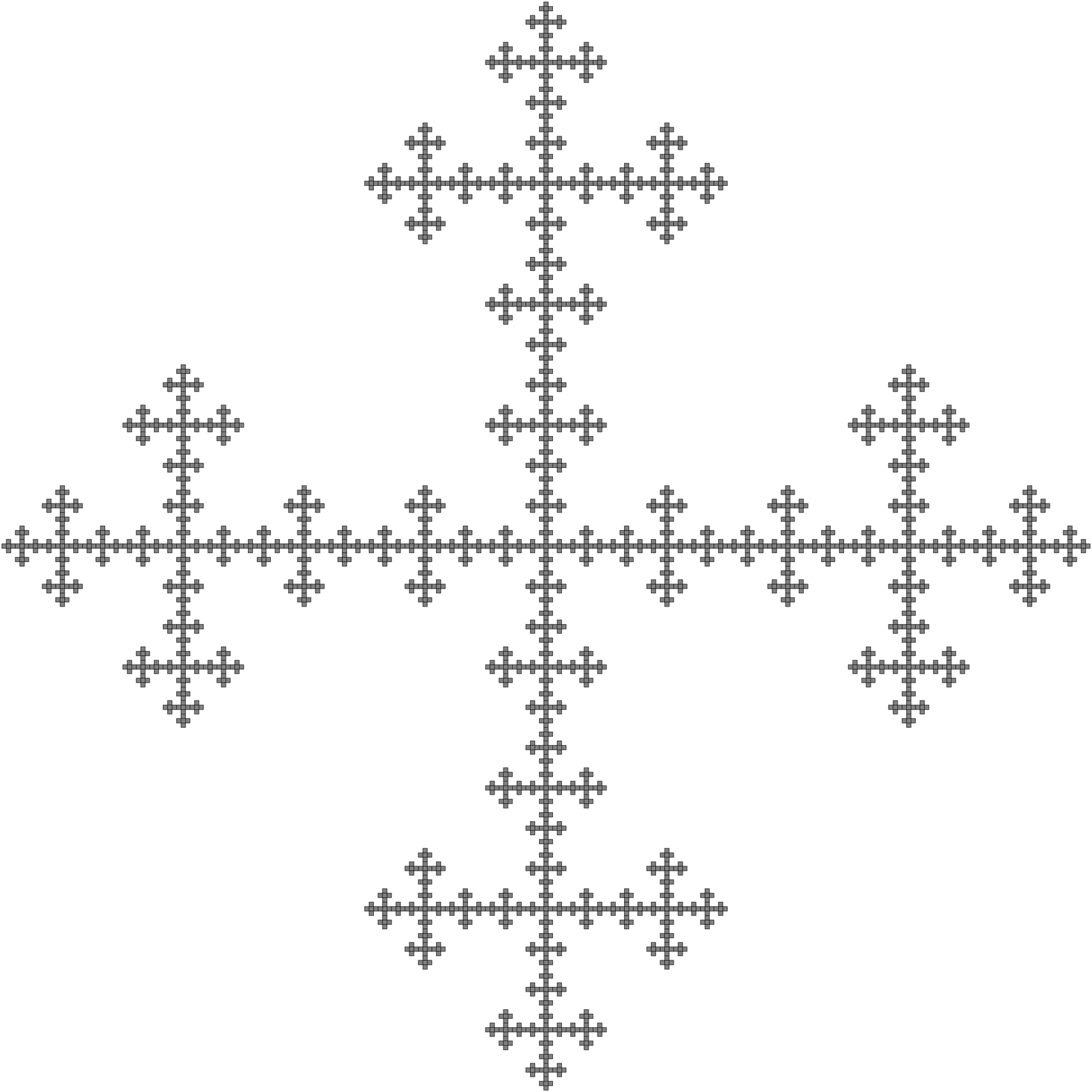}} & \raisebox{-\totalheight}{\includegraphics[scale=0.5] {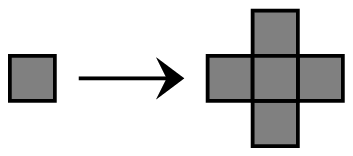}} & \ \newline $\frac{\log(5)}{\log(3)} \approx 1.46$\\
    \hline
    Empty-Bottles (Custom) & \raisebox{-\totalheight}{\includegraphics[scale=0.03] {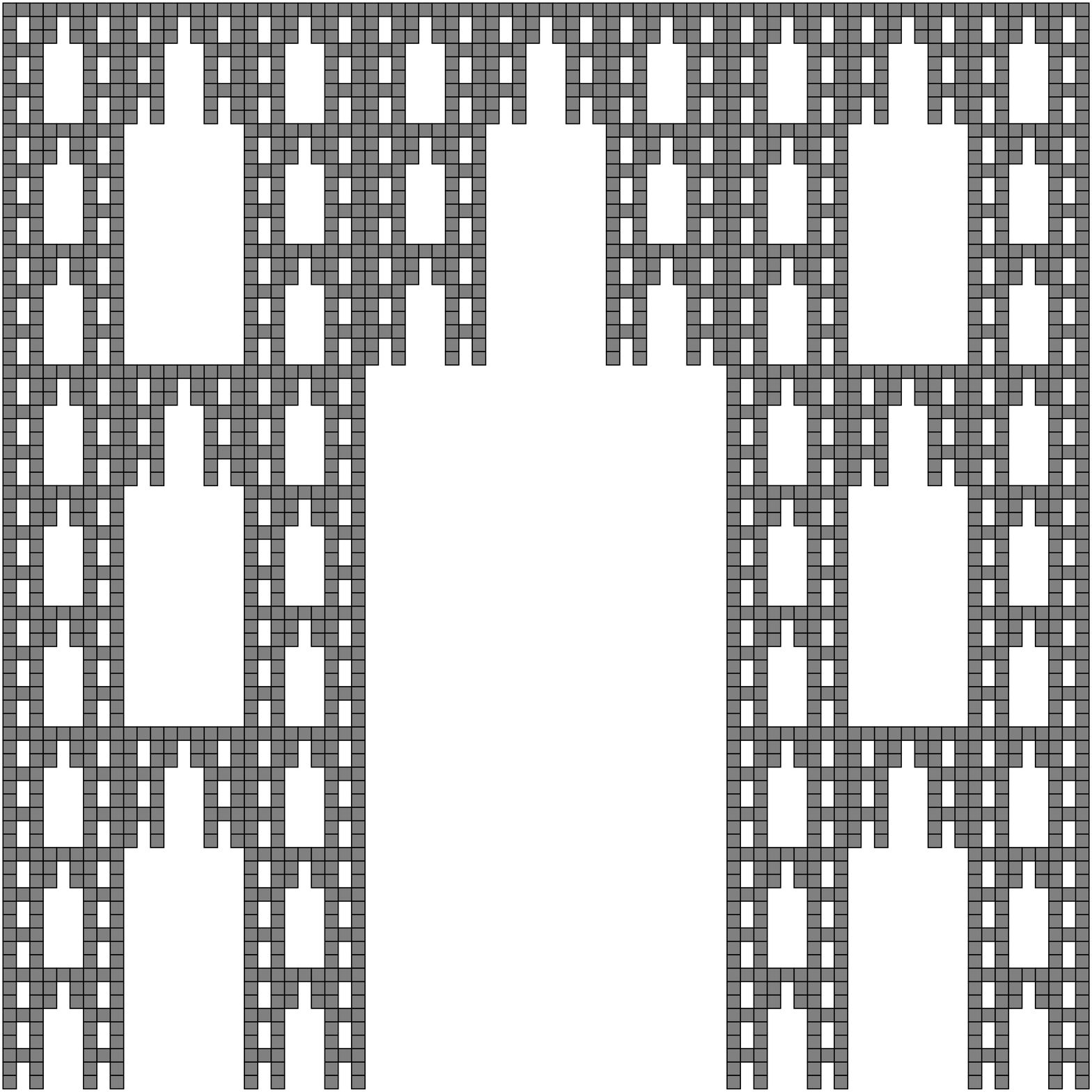}} & \raisebox{-\totalheight}{\includegraphics[scale=0.5] {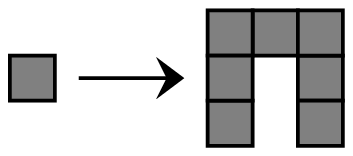}} & \ \newline $\frac{\log(7)}{\log(3)} \approx 1.77$\\
    \hline
    Cantor set & \raisebox{-\totalheight}{\includegraphics[scale=0.010] {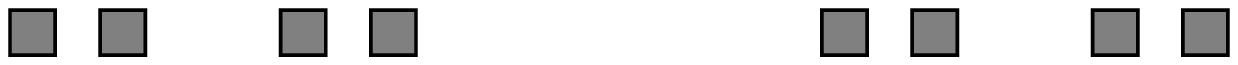}} & \raisebox{-\totalheight}{\includegraphics[scale=0.5] {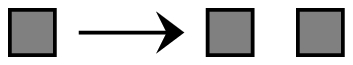}} & \ \newline $\frac{\log(2)}{\log(3)} \approx 0.63$\\
    \hline
    \end{tabular}
    \caption{Example fractals of the NBB family.}
    \label{tbl:discrete-fractals}
    \end{center}
\end{table}

Given that $k$ is fixed, and $n$ scales up by factors of $s$ as the fractal level increases, the space used by a fractal, denoted as $\mathcal{V}(\fracnot{n}{k}{s})$ may be expressed as:
\begin{equation}
    \mathcal{V}(\fracnot{n}{k}{s}) = k^r
\end{equation}
where $r = \log_{s}(n)$ is defined as the scale level.
\subsection{The two problems with the embedded representation}
In the embedded representation, as the fractal gets larger, 
the number of fractal data-elements will become asymptotically smaller than the number of non-fractal elements (the empty spaces of the embedding), bringing up two problems:
\begin{itemize}
    \item Problem 1 (P1): If treated as a default GPU program, the number of computational resources (threads) mapped to the problem will grow in terms of the bounding-box (embedding space), and not in terms of the number of elements of the fractal which is what is actually needed.
    \item Problem 2 (P2): The memory usage will increase in terms of the bounding-box, and not in terms of the fractal, putting a very early limit on the largest problem size that fits in the GPU.
\end{itemize}

\begin{figure}[ht!]
\centering
\includegraphics[scale=0.4]{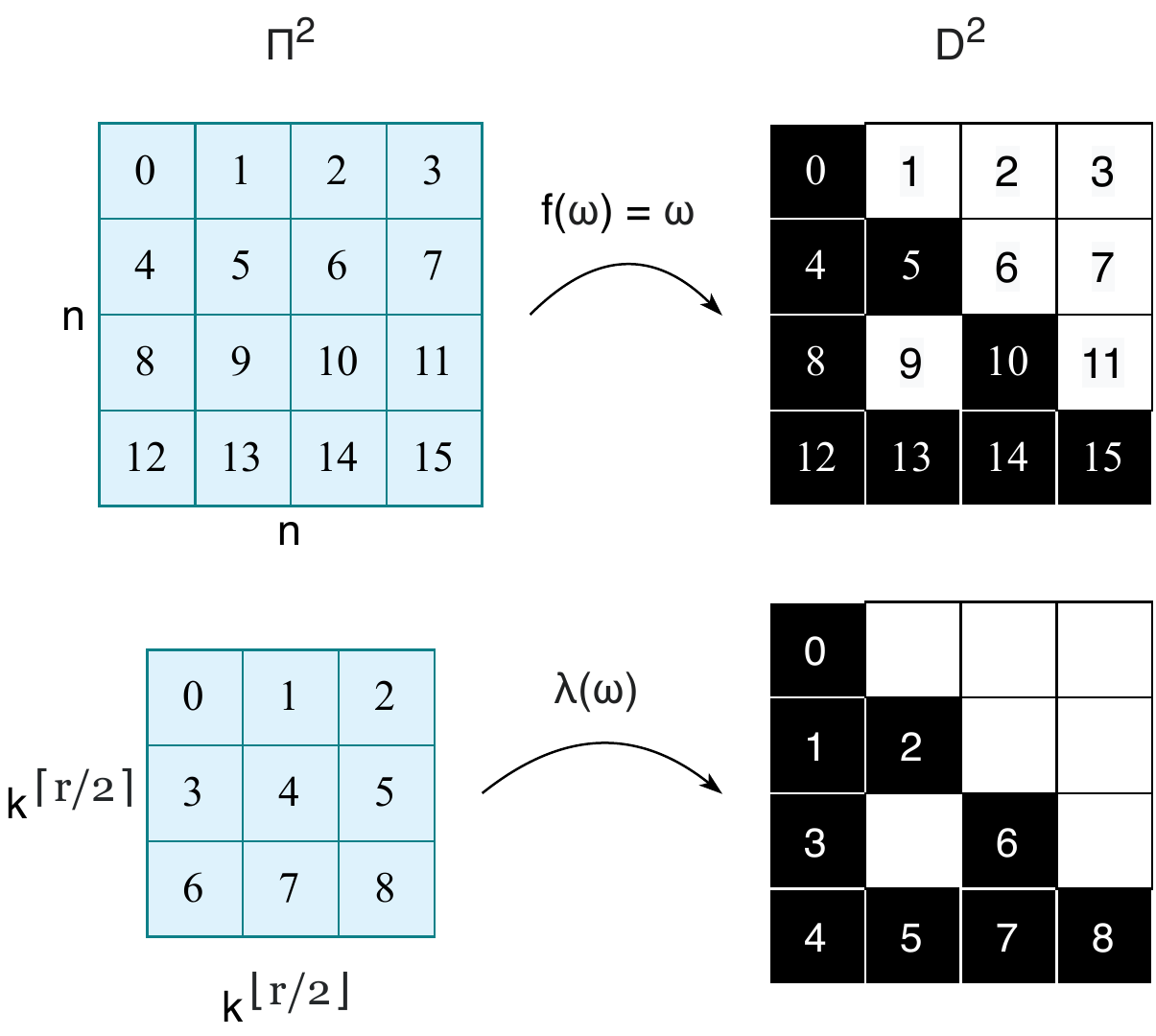}
\caption{Mapping of threads (left) to data (right) using BB (top) and $\lambda(\omega)$ (bottom).}
\label{fig:bb-vs-lambda}
\end{figure}

Navarro \textit{et al.} proposed in 2020 an efficient GPU Tensor-core accelerated thread map for NBB fractals, denoted $\lambda(\omega)$, that allows using just the necessary number of threads to reach the fractal data elements in the embedding representation. In terms of CUDA programming, the approach by Navarro \textit{et al.} compacts the CUDA Grid of threadblocks to the minimum and necessary for fractal, leading to a significant speedup in performance and thus accelerating applications that act on the fractal. Figure \ref{fig:bb-vs-lambda} illustrates the benefits of using $\lambda(\omega)$ compared to a bounding-box (BB) approach.

\begin{figure}[ht!]
\centering
\includegraphics[scale=0.7]{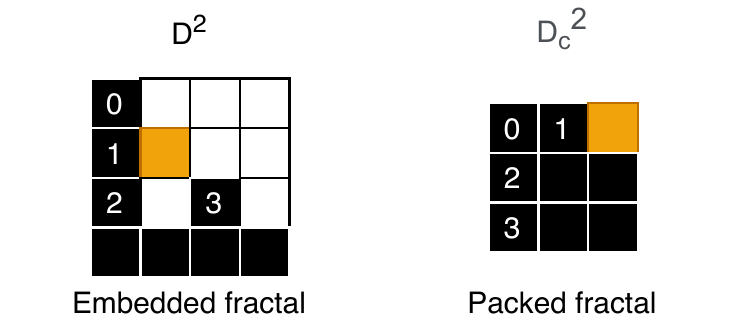}
\caption{Neighborhood layouts for the orange cell when using embedded and compact representations.}
\label{fig:neighbor}
\end{figure}

The approach with $\lambda(\omega)$ solves problem P1 and improves performance significantly, and can also solve problem P2 for computations that do not require exploring neighborhood, by using a compact representation of the fractal. However, $\lambda(\omega)$ alone cannot solve problem P2 for stencil-like computations because neighbors follow a layout that is different from the embedded one (see Figure \ref{fig:neighbor}).
Applications such as FDTD PDE solvers, cellular-automata/spin-models simulation and signal processing among others rely on neighboring cells to compute a cell state. 

This technical report presents a Tensor Core GPU approach capable of doing parallel computations, including stencil-like ones, on NBB fractals using their compact representation. Experiments using a NVIDIA A100 GPU show that the approach is up to $11\times$ faster and $234\times$ more memory efficient than a bounding box approach. The remaining sections cover related work, the proposed approach, results and conclusions.

\section{Related Work}
GPU mapping techniques started in 2008 for mapping threads onto triangular spaces such as lower-triangular matrices. Jung \textit{et al.} \cite{Jung2008} developed in 2008 an algorithm to map triangular (2-simplex) shaped data to a rectangular box which significantly reduces memory requirements in applications like LU or Cholesky decomposition. In particular, the total memory used is reduced in half. Ries \textit{et al.} \cite{Ries:2009:TMI:1654059.1654069} developed in 2009 a method to compute the inverse of triangular matrices by developing a recursive parallel space mapping from a compact rectangular domain using GPU. Navarro \textit{et al.} proposed in 2014 a GPU block-space mapping for 2-simplex and 3-simplex shaped data \cite{DBLP:conf/hpcc/NavarroH14,CLEI-2016-navarro,navarro2018competitiveness}. 
Navarro \textit{et al.} \cite{8291959,NAVARRO2020158} expanded the idea of GPU thread mapping for fractal domains, by proposing the $\lambda(\omega)$ map for NBB fractals.  

Regarding the use of Tensor Cores in non-Machine-Learning tasks, Carrasco \textit{et al.} \cite{inproceedingsTC} studied the potential speedup of using tensor cores on arithmetic reductions. The authors concluded that the tensor core based reduction was indeed theoretically faster. Navarro \textit{et al.} \cite{9147055} extended the work and proposed a chained tensor core version that achieved $3.2\times$ over a traditional CUDA implementation. Dakkak \textit{et al.} \cite{10.1145/3330345.3331057} also proposed and arithmetic reduction approach using tensor cores, as well as a parallel scan approach, reporting significant speedups. 

The recent use of tensor core units (TCU) for non-Machine-Learning applications combined with the opportunity to solve problem P2 for stencil-like computation puts a great motivation to research on new Tensor Core GPU techniques to handle compact fractals efficiently.

\section{A Tensor Core Approach for Fractals in Compact Representation}
\begin{figure}[ht!]
\centering
\includegraphics[scale=0.9]{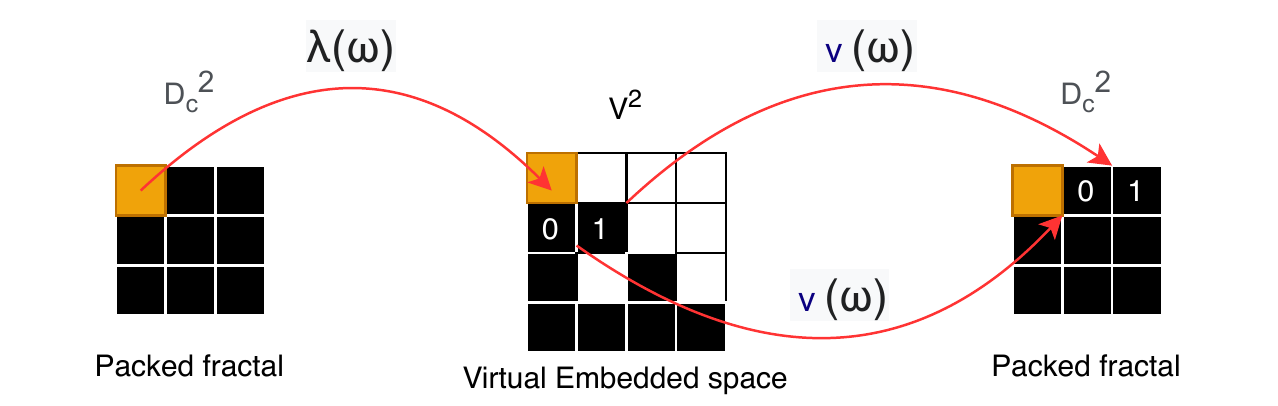}
\caption{Example of the process done to read values from neighboring cells. The orange cell represents the actual element to simulate. To identify the neighbors, it has to be mapped to a virtual embedded space.}
\label{fig:example-nu}
\end{figure}

The new approach consists of combining $\lambda(\omega)$ with a new proposed map, denoted $\nu(\omega)$, in such a way that the former acts as a tensor-core-accelerated function from compact-space to embedded-space, while the latter acts as a tensor-core-accelerated function from embedded-space to compact space. By using the two maps, it is possible for all fractal locations to explore their neighborhoods without needing to expand the fractal into embedded space. 
Figure \ref{fig:example-nu} illustrates how the approach uses both maps in conjunction.

\begin{figure}[ht!]
\centering
    \includegraphics[scale=0.7]{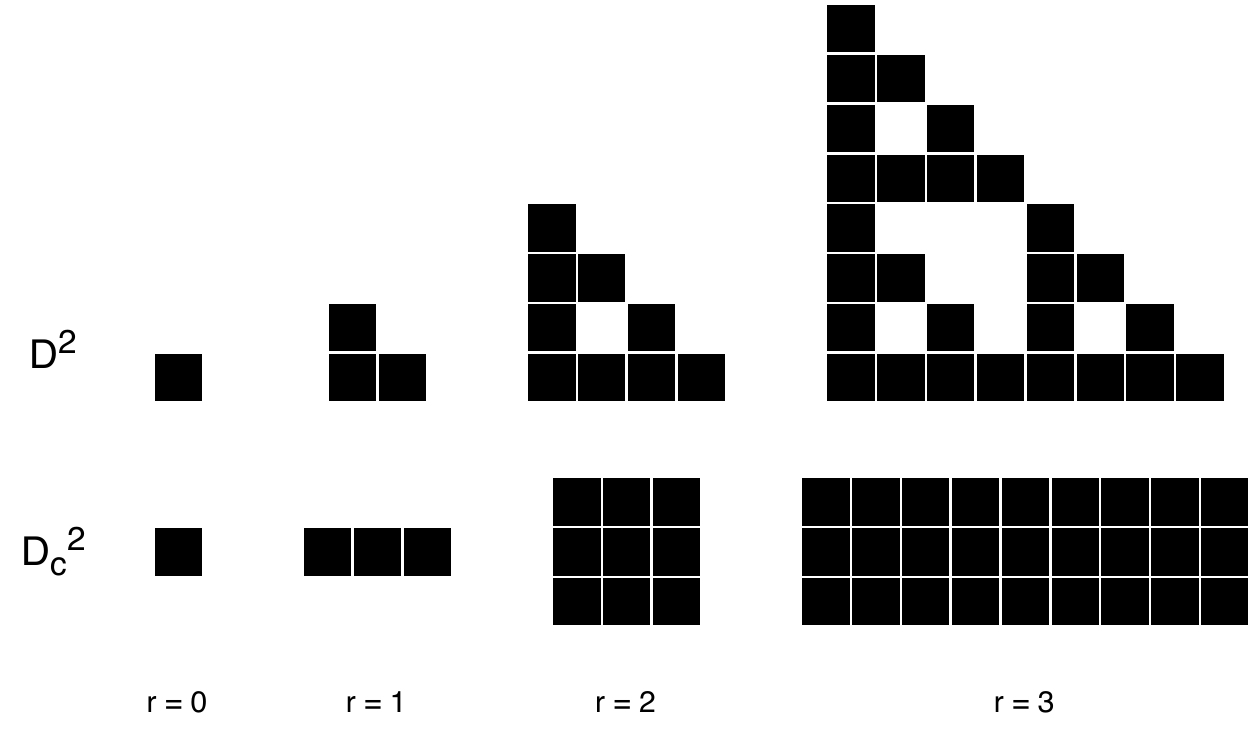}
\caption{The Sierpiński Triangle (top) and its compact form (bottom) at each scale level. Note that the compact version is constructed with an unfolding process.}
\label{fig:compact}
\end{figure}

The formulation of $\nu(\omega)$ follows an approach similar to the one made for $\lambda(\omega)$ \cite{NAVARRO2020158}. $\nu(\omega)$ is a space transformation that receives a coordinate of an element in embedded fractal space $\mathbf{D}^2$ and returns its corresponding coordinate in compact space $\mathbf{D}_c^2$. In principle, $\nu(\omega)$ acts as a inverse to $\lambda(\omega)$. Both spaces start with the coordinate $(0,0)$ in the upper-left corner, increasing $x$ to the right and $y$ downwards. The intuition behind the compact space layout is similar to an unfolding process. Starting with $r=0$, the compact space contains one element. From that point and on, it replicates itself $k$ times to the right when $r$ is odd, and down when even. Figure \ref{fig:compact} illustrates the process level-by-level.

\subsection{Formulation of $\nu(\omega)$}
Let $H(x, y, \mu)$ denote a helper function that takes coordinates $x$ and $y$ and the current scale level of the fractal and returns a unique ID of the replica that coordinate belongs to. The ID must be a unique number from $0$ to $k-1$. $H(x, y, \mu)$ depends on the particular fractal to map and can be a look-up table of size $k$ or a procedural function, \textit{e.g.} for the Sierpiński triangle, $H(x, y, \mu)$ may be a procedural function defined as:
\begin{equation} \label{eq:H}
H(x, y, \mu) = \left\lfloor \frac{x\ \mathbf{mod}\ 2^{\mu+1}}{2^\mu}\right \rfloor+ \left\lfloor \frac{y  \ \mathbf{mod}\ 2^{\mu+1}}{2^\mu}\right \rfloor
\end{equation}
with $\mathbf{mod}$ to denote the modulus operator.
It is important to note that the ID of the replica has to be in correspondence with the numbering of the replica used in $\lambda(\omega)$.

Let $\tau(k, \mu)$ denote another function that receives the numbers of replicas at the next scale level $k$ and the scale level $\mu$. It returns the offset of the replica at a particular scale level. $\tau(k, \mu)$ is defined as:

\begin{equation} \label{eq:t}
\tau(k, \mu) = (\tau_x(k, \mu) , \tau_y(k, \mu))
\end{equation}
 with
\begin{equation} \label{eq:tx}
\tau_x(k, \mu) = k^{\frac{\mu}{2}} ((\mu+1)\ \mathbf{mod}\ 2)
\end{equation}
\begin{equation} \label{eq:ty}
\tau_y(k, \mu) = k^{\frac{\mu}{2}} (\mu\ \mathbf{mod}\ 2) \space \space
\end{equation}
for $x$ and $y$ coordinates. The specific coordinate in the compact space is:
\begin{equation} \label{eq:realnu}
\nu(\omega) = (\nu_x(\omega), \nu_y(\omega))
\end{equation}

with
\begin{equation} \label{eq:nux}
\nu_x(\omega) = \sum_{\mu=0}^{r-1}\tau_x H(\omega_x,\omega_y,\mu) \\
\end{equation}

\begin{equation} \label{eq:nuy}
\nu_y(\omega) = \sum_{\mu=0}^{r-1}\tau_y H(\omega_x,\omega_y,\mu)
\end{equation}

Although $\nu(\omega)$ and $\lambda(\omega)$ have an efficient upper bound of $\mathcal{O}(\log_2(\log_s(n)))$ by using a parallel reduction with the threads contained in the $\omega$ GPU block \cite{NAVARRO2020158}, it can be further accelerated with tensor core units (TCU). The acceleration with tensor cores is achieved by encoding sum of products from Eqs. \ref{eq:nux}, \ref{eq:nuy} as matrix-multiply-accumulate (MMA) operations as shown in Figure \ref{fig:tensor_map}. 

\begin{figure}[ht!]
    \centering
        $ A=
        \begin{pmatrix}
        \tau_x(k, 0) & \tau_x(k, 1) & \dots & \tau_x(k, \mu-1)\\
        \tau_y(k, 0) & \tau_y(k, 1) & \dots & \tau_y(k, \mu-1)\\
        0 & 0 & \dots & 0\\
        \vdots & \vdots & \ddots & \vdots\\
        0 & 0 & \dots & 0\\
        \end{pmatrix}
        $
        $B=
        \begin{pmatrix}
        \mathbf{H}(x,y,0) & 0 & \dots & 0\\
        \mathbf{H}(x,y,1) & 0 & \dots & 0\\
        \mathbf{H}(x,y,2) & 0 & \dots & 0\\
        \vdots & \vdots & \vdots & \ddots & \vdots\\
        \mathbf{H}(x,y,\mu-1) & 0 & \dots & 0\\
        \end{pmatrix}
$
        \caption{The encoding of $\nu(\omega)$ into a typical MMA operation. Note that these matrices sizes are $\mu \times \mu$. In practice these are then loaded into a $16\times16$ tensor core fragment.}
    \label{fig:tensor_map}
\end{figure}

These MMA operations are assigned one per warp, and given that the MMA can be perceived as a $O(1)$ operation (hardware level), potential performance speedup is expected. The programming of tensor cores was done using CUDA's  WMMA API available. It worth noticing that a tensor Core fragment can take different sizes, depending on the data type. This work chose FP16 matrix multiplication with FP32 accumulation of fragments of size $16\times16$. 

\begin{figure}[ht!]
\centering
\includegraphics[scale=0.7]{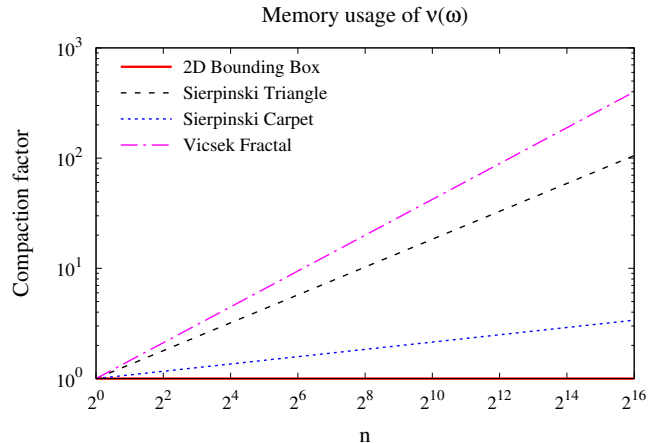}
\caption{The compaction factor for a few NBB fractals, higher is better.}
\label{plot:theoretical}
\end{figure}

Figure \ref{plot:theoretical} shows the compression factor achieved for three different NBB fractals when using their compact form instead of the embedded one. As one can note, this compaction factor grows exponentially as the size of the fractal increases. At $n=2^{16}$ the compact version of the Sierpiński triangle uses $\sim1\%$ the space otherwise used by a BB approach. This enables the possibility to process large fractal sizes on a single GPU that were not possible before.

For an application to successfully compute a stencil-like computation using its neighborhood, the following steps would be employed with this approach. First, parallel resources are mapped to compact parallel space, where they can easily access their corresponding data element in memory. In order to access nearest neighbors, it computes its corresponding coordinate in virtual embedded space using $\lambda(\omega)$ so it can know the neighbor's coordinates (\textit{i.e.}, $(x+1,y), (x-1,y), (x,y+1), (x,y-1))$. Then it maps the neighbor's coordinates back to compact space using $\nu(\omega)$. Once the neighborhood is known in compact space, the parallel resource can do the application's stencil-like computation as usual. 

\begin{figure}[ht!]
\centering
\includegraphics[scale=0.5]{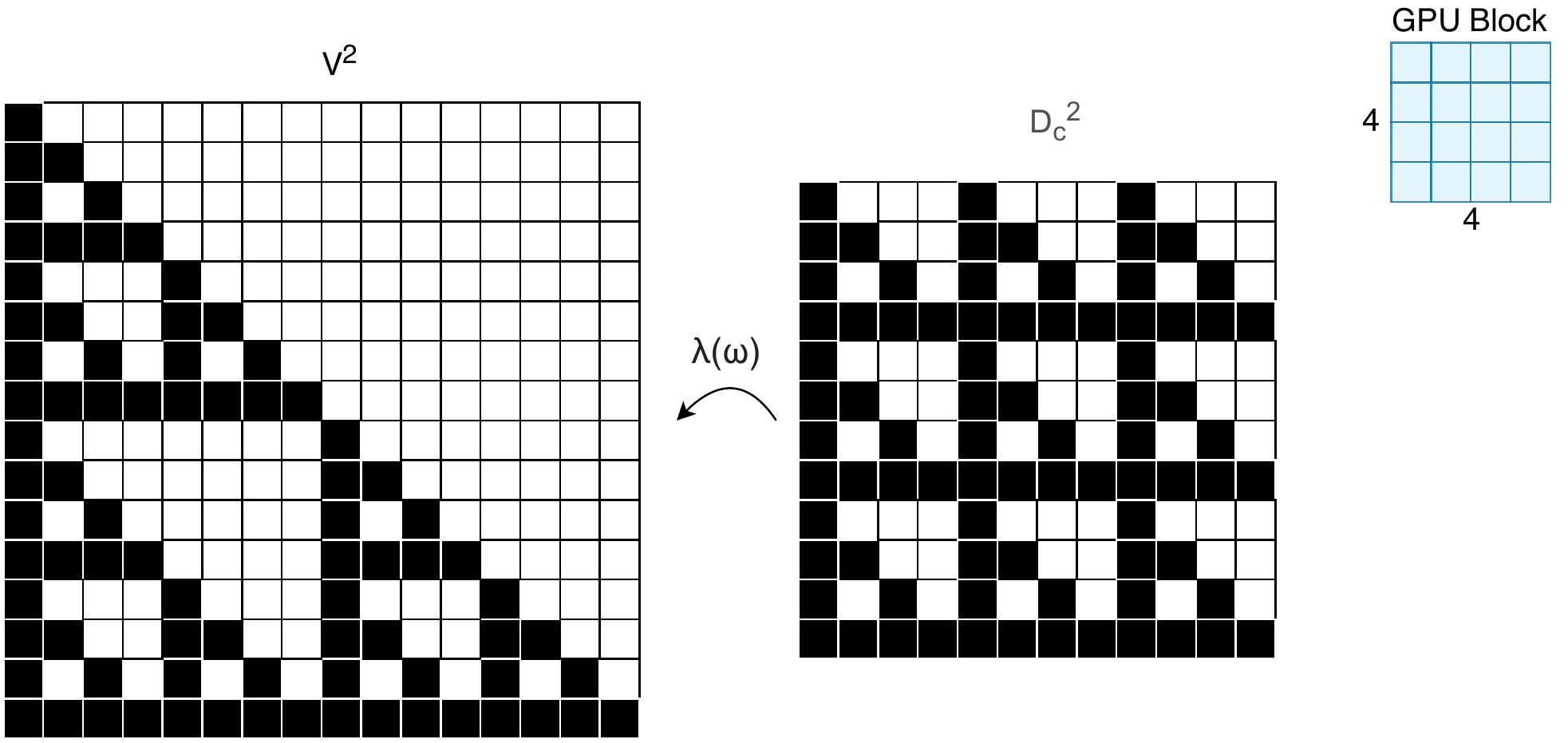}
\caption{Example of how the fractal is saved in memory using a blocked approach. Here, the GPU block size is $4\times4$ and the compact space is a collection of small bounding-boxes.}
\label{fig:coarser}
\end{figure}

To make the process even more efficient on GPU, one can use a blocked approach in which one block of $\rho \times \rho$ threads are grouped as one coarse-coordinate to map. This means that we map a coarser version of the fractal and fewer maps are performed, decreasing execution divergence and bringing up the possibility to cooperatively calculate the map expression by all threads of the block. In the compact space, blocks would be tightly packed, and inside each block one would find a miniaturized version of the embedded fractal that acts as a micro-brute-force solution. Although this miniature fractals introduce an extra memory usage and may introduce some unused threads, the overhead is constant and becomes less significant as the fractal becomes larger. Figure \ref{fig:coarser} shows an example of the packing done using a blocked approach.

\section{Experimental Results}
Three different approaches were put to test in terms of performance,
\begin{enumerate}
    \item Bounding-box (BB)
    \item Compact-grid with expanded embedded fractal (referred to as $\lambda(\omega)$)
    \item Compact-grid and compact-fractal (referred to as $\nu(\omega)$). 
\end{enumerate}
With the third one being the proposed approach of this work.
The chosen test benchmark was Conway's game of life running on the Sierpinski Triangle, with an adapted Moore's neighborhood (with adapted life/death conditions as well). The three methods were tested using the setups listed in Table \ref{table_hardware}. 
\begin{table}[ht!]
\caption{Hardware setups used.}
\begin{center}
\begin{tabular}{| c | c | l |}
\hline
Setup & Device	&	Model\\
\hline
  & GPU	&	Titan V, 5120 CUDA cores 12GB \\
0 & CPU	&	Intel i7-6950X 10-core Broadwell \\
  & RAM	&	128GB DDR4 2400MHz\\
  & TCU & 640 Tensor Cores First Gen \\
\hline
  & GPU	&	A100, 6912 CUDA cores, 40GB  \\
1 & CPU	&	Dual AMD Rome 7742, 128 cores total \\
  & RAM	&	1 TB DDR4\\
    & TCU & 432 Tensor Cores Third Gen \\
\hline
\end{tabular}
\end{center}
\label{table_hardware}
\end{table}

The performance metric is the average execution time of $100$ runs of $1000$ iterations for each system and each method. Additionally, the Speedup metric is used to compare the benefit of each method. The Speedup is defined as
\begin{equation}
    S=\frac{T_{ref}}{T_{comp}}
\end{equation}
where $T_{ref}$ is the bounding-box running time, and $T_{comp}$ is the time of the second or third approach. The level of the fractal was varied from from $r=1$ to $r=16$ and the size of the block of threads $\rho \times \rho$ from $\rho=2$ to $\rho=32$.
Additionally, two versions of each method were tested, with and without the use of shared memory. The results only use the fastest version of each one. For BB and $\lambda(\omega)$ this was without shared memory, and $\nu(\omega)$ with shared memory.


\subsection{Case study: Sierpiński Triangle}
\begin{figure*}[ht!]
\begin{center}
\includegraphics[scale=0.50]{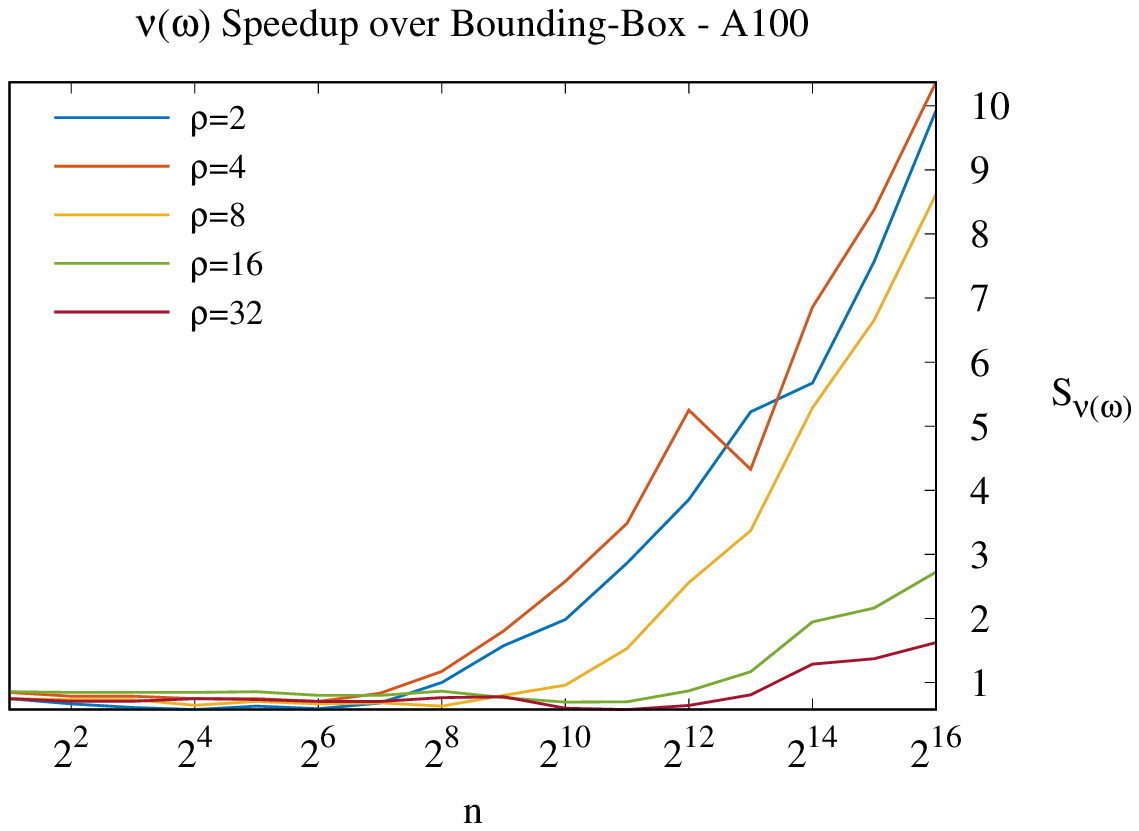}
\includegraphics[scale=0.50]{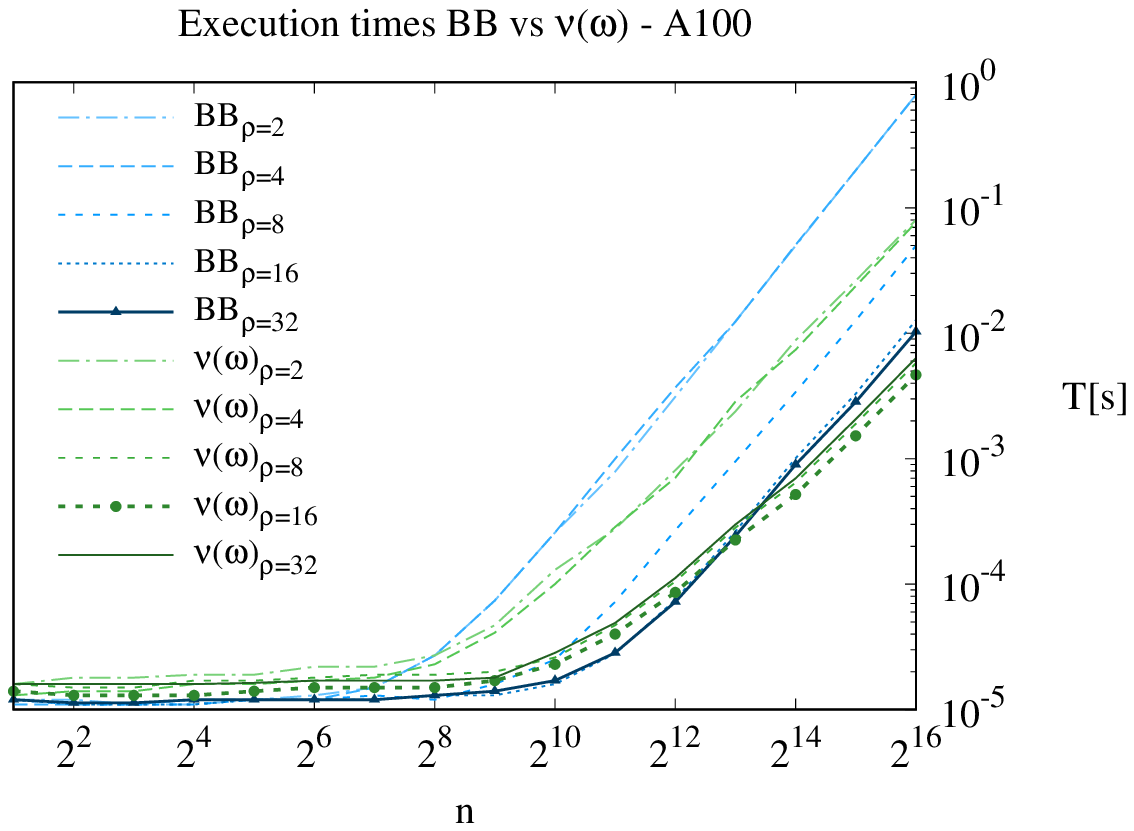}

\caption{On the left the Speedups of $\nu(\omega)$ compared to BB. And on the right, the execution times per iteration of $\nu(\omega)$ and BB.}
\end{center}
\label{fig:speedup_times}
\end{figure*}

Figure \ref{fig:speedup_times} shows the Speedup and execution times of BB and $\nu(\omega)$. $\lambda(\omega)$ was omitted from these results as it represents a lower bound in execution time, as presented later.
Speedup curves show that using a compact version of the fractal in sizes below $n=2^7$ doesn't provide much difference in performance, with a small penalty due to the overhead of the $l+1$ mapping. As $n$ grows further 2 things happen: (1) for block sizes $2$, $4$ and $8$ the Speedup starts to increase with $n$ reaching a top Speedup of $\sim10.2$. (2) For blocks sizes $16$ and $32$ the speedup enters a valley below $1$, until it increases again reaching up to $\sim2.8$ of Speedup. The execution times provide further insight showing that the best version of BB is with $\rho=32$ and the best version of $\nu(\omega)$ is when $\rho=16$. Comparing the best version of each, we can note that starting from $n\ge2^{13}$, $\nu(\omega)$ becomes the better option.

\begin{figure*}[ht!]
\begin{center}
\includegraphics[scale=0.50]{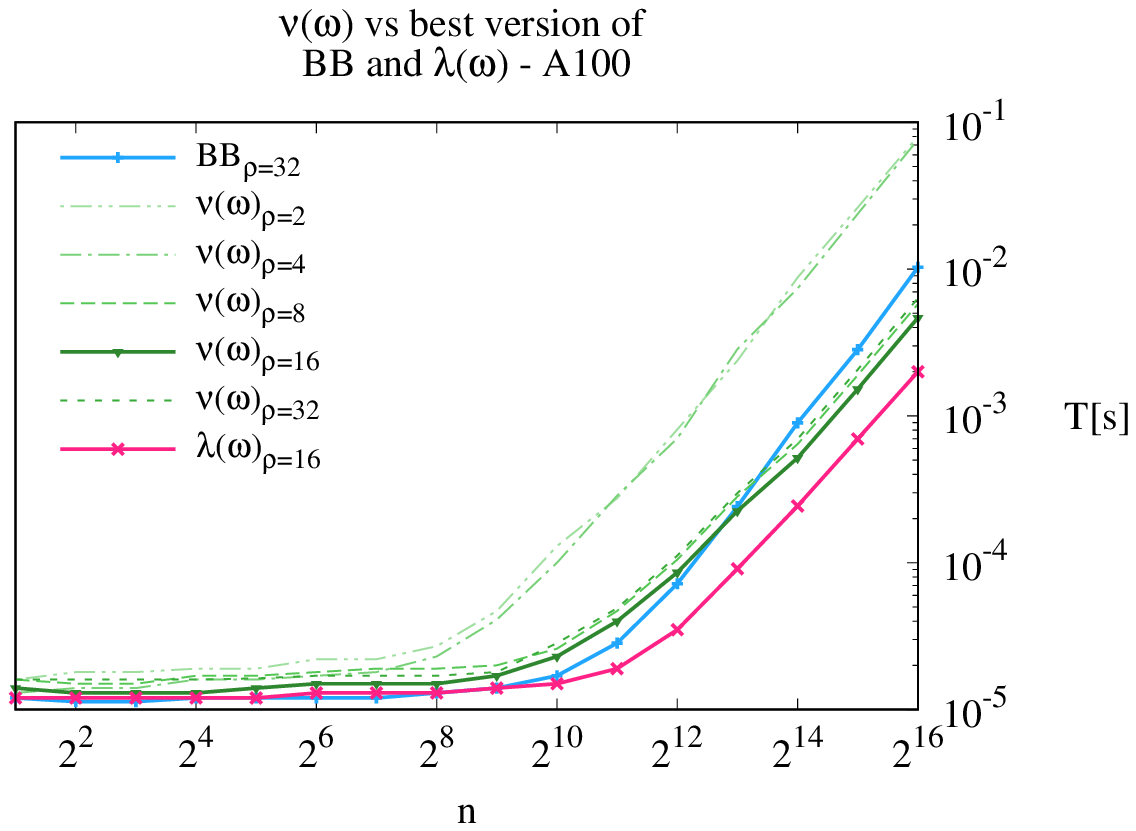}
\includegraphics[scale=0.50]{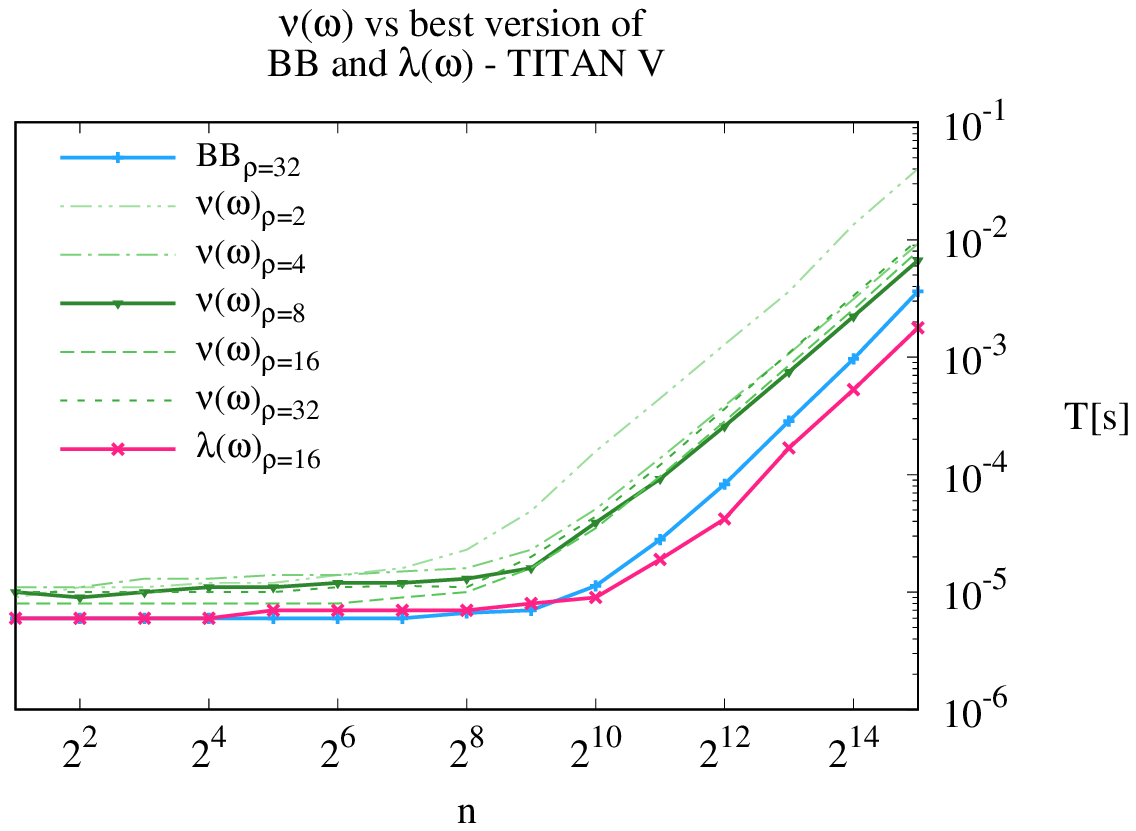}
\caption{Plots showing a comparison of the best running times of BB and $\lambda(\omega)$ versus the running times of $\nu(\omega)$ in all configurations.
}
\end{center}
\label{fig:best}
\end{figure*}

The Figure \ref{fig:best} compares the execution times of $\nu(\omega)$ and the best versions of BB and $\lambda(\omega)$. 

Results with A100 shows that from $n\ge2^{13}$, $\nu(\omega)$ with $\rho=8, 16, 32$ is faster than the best version of BB. And only $\rho=2, 4$ is slower than BB. On the other hand, results in the Titan V shows that not even the best version of $\nu(\omega)$ is faster than BB. In both GPUs the execution time of BB is similar, but the mappings runs significantly slower in the Titan V.

\begin{figure*}[ht!]
\begin{center}
\includegraphics[scale=0.50]{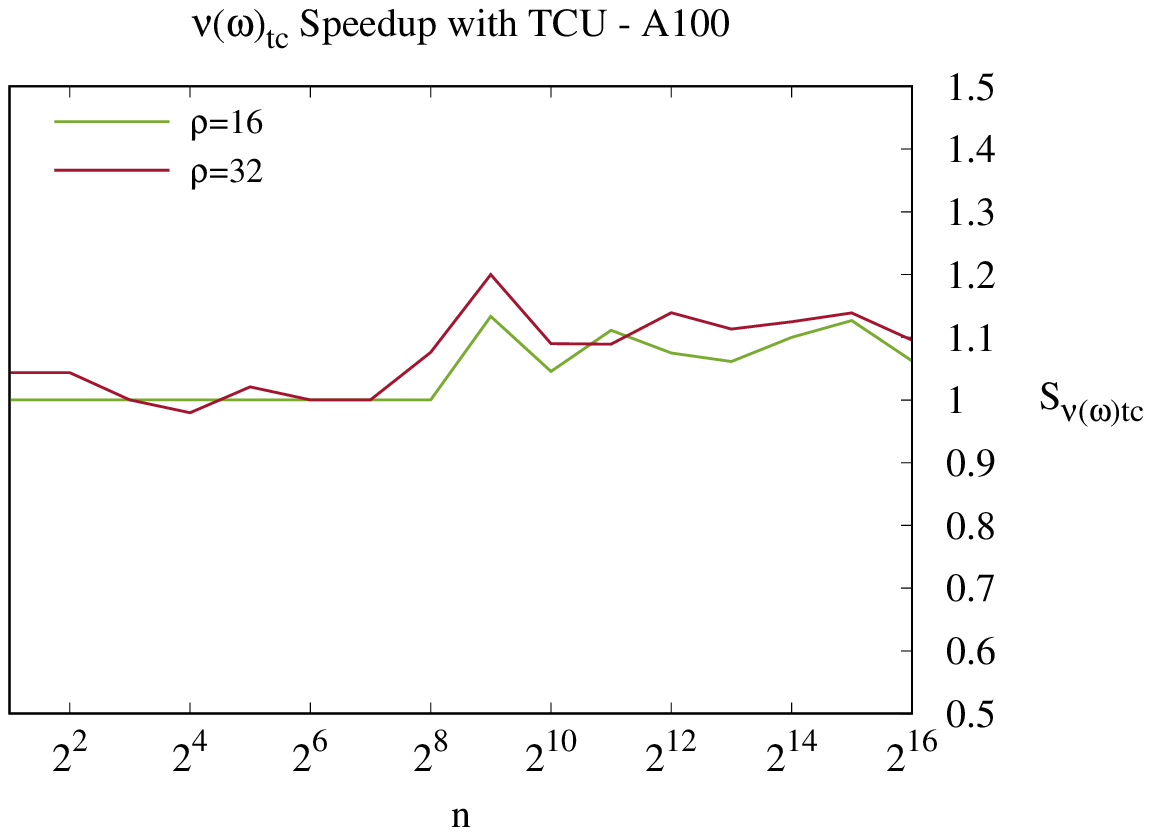}
\includegraphics[scale=0.50]{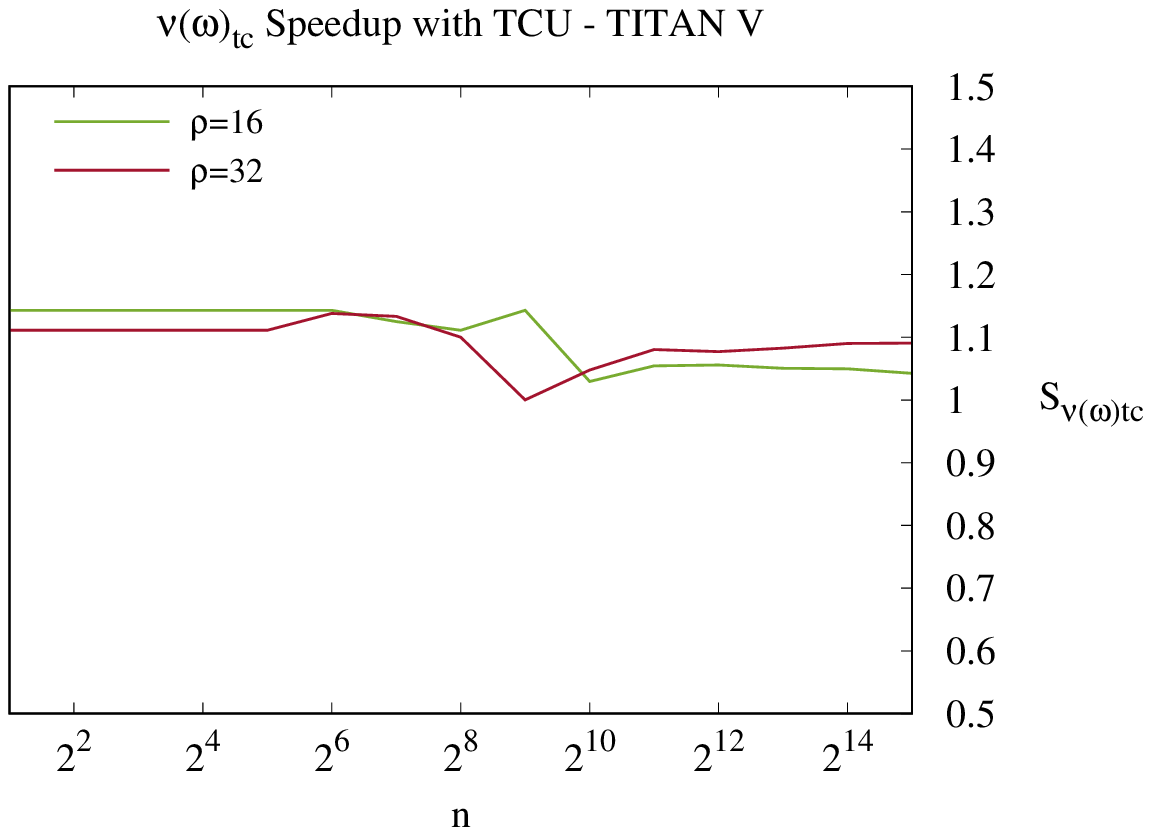}
\caption{Speedup of the tensor core version of $\nu(\omega)$ with respect to the non-tensor core version in all three generations of Tensor Cores (Volta, Turing and Ampere)}
\end{center}
\label{fig:tensor_perf}
\end{figure*}

In terms of TCU, Figure \ref{fig:tensor_perf} shows the extra performance by using tensor cores. The tests contains the behavior using first and third generation tensor core units. Starting with the A100 GPU, the TCU provide up to $10\%$ of extra performance once $n\ge2^8$. In the Titan V, the performance from TCU is maintained across values of $n$, with again, at most ~$10\%$ of extra performance. 

\begin{figure}[ht!]
\centering
\includegraphics[scale=0.50]{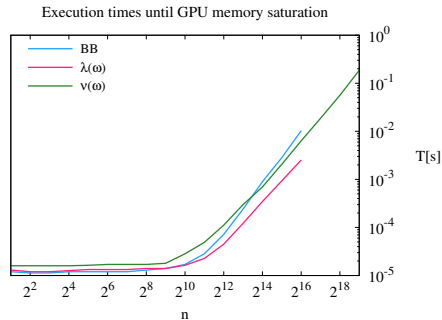}
\caption{A comparison of how the compact approach can reach much larger sizes on a NVIDIA A100 GPU.}
\label{fig:sizes_comp}
\end{figure}

Finally, Figure \ref{fig:sizes_comp} serves as an illustration of how the full compact approach is capable of going beyond $n=2^{16}$, up to $n=2^{20}$ in a single A100 GPU. At each size increment there is a higher speedup, showing evidence that at larger problem sizes the compact strategy approaches the performance performance of the expanded $\lambda(\omega)$ approach, while keeping the benefits of using a compact representation.

\section{Discussion and Conclusions}
This work presented a GPU approach for handling discrete fractals efficiently on GPUs. By using their compact representation, two benefits occur: i) computation is employed only in the fractal elements and not in empty embedding space, bringing significant speedup when compared to a bounding-box approach, and ii) it produces an exponential reduction in memory usage, in proportion to the Hausdorff dimensions of the fractal. These aspects bring the possibility to process fractals of much larger size than before.
Tests with the A100 GPU allowed memory saving of up to $247\times$. This enabled the possibility to increase the fractal scale from $15$ to $20$, which equals managing problem sizes up to $32\times$ larger than scale $15$. 
In terms of performance, results using the Sierpiński Triangle showed up to $11\times$ of speedup compared to a Bounding Box approach. It is worth noticing that the performance and memory optimizations keep increasing with $n$, therefore future generation of GPUs can report even better results, and reach problem sizes that would be infeasible with traditional approaches that use the fractal's expanded embedded form. 
Adapting the computation of $\lambda(\omega)$ and $\nu(\omega)$ to tensor core MMAs provided a $10\%$ of extra performance compared to just using regular CUDA cores. Future improvements to tensor cores, both in quantity and performance, would provide an even better performance improvement. As future work, it would be useful to come up with a way to build arbitrary fractal structures by combining different NBB fractals at each scale level. As a final conclusion, the proposed approach can be adopted by the community to accelerate simulations on NBB fractals and understand phenomena at larger scales.

\section*{Acknowledgement}
This research was supported by the Temporal research group (\url{http://temporal.uach.cl}), ANID Fondecyt Iniciacion grant \#11180881 and the Patagón supercomputer of Universidad Austral de Chile (FONDEQUIP EQM180042).
\bibliographystyle{splncs04}
\bibliography{main}
\end{document}